\documentclass[10pt,aps,prd,superscriptaddress,nofootinbib,nobibnotes,longbibliography,floatfix,twocolumn]{revtex4-2}

\usepackage{bm}
\usepackage{mathtools,
amsmath,
amssymb,
amsfonts,
mathrsfs,
chngcntr,
multirow}

\let\cc\corresponds
\let\corresponds\relax
\usepackage{mathabx}
\let\corresponds\cc

\usepackage{mathbbol}

\usepackage[utf8]{inputenc}
\usepackage[T1]{fontenc}

\usepackage{soul}

\usepackage[dvipsnames]{xcolor}
\usepackage[unicode]{hyperref}
\hypersetup{colorlinks=true, citecolor=MidnightBlue,
            linkcolor=MidnightBlue, urlcolor=MidnightBlue, linktocpage=true}
\usepackage[normalem]{ulem}

\usepackage{graphicx}

\begin{document}

\title{A parity selection rule for regular black holes}

\author{Farid Thaalba}
\affiliation{SISSA, Via Bonomea 265, 34136 Trieste, Italy}
\affiliation{INFN, Sezione di Trieste, Via Valerio 2, 34127 Trieste, Italy}

\author{Julio Arrechea}
\affiliation{SISSA, Via Bonomea 265, 34136 Trieste, Italy}
\affiliation{INFN, Sezione di Trieste, Via Valerio 2, 34127 Trieste, Italy}
\affiliation{IFPU, via Beirut 2, 34014 Trieste, Italy.}

\author{Stefano Liberati}
\affiliation{SISSA, Via Bonomea 265, 34136 Trieste, Italy}
\affiliation{INFN, Sezione di Trieste, Via Valerio 2, 34127 Trieste, Italy}
\affiliation{IFPU, via Beirut 2, 34014 Trieste, Italy.}

\begin{abstract}
Regular black hole metrics are usually studied kinematically, but a finite-curvature static core does not guarantee that the underlying theory can consistently evolve generic matter through a regular center. We derive a necessary local consistency condition within the most general class of action-based, identically conserved, second-order gravitational field equations in spherical symmetry. Regularity requires that the two functions defining the theory have opposite parities under reversal of the signed radial coordinate, together with additional center-regularity and nondegeneracy conditions. In the integrable sector, this criterion is equivalent to requiring the generalized Misner--Sharp--Hernandez mass to be odd across the center, to vanish cubically there, and to contain no point-mass contribution. For theories reconstructed from static one-parameter vacuum families, the condition becomes covariance under simultaneous reversal of radius and mass. The theories associated with the Hayward and Dymnikova geometries satisfy this selection rule. In contrast, the Bardeen theory does not, demonstrating that curvature regularity of a static solution is insufficient for dynamical consistency with generic matter. We also characterize an infinite class of admissible theories containing Hayward-like black holes with de Sitter cores. The selection rule provides a necessary condition for theories intended to describe regular collapse, but does not by itself establish well-posedness or guarantee a nonsingular endpoint.

\end{abstract}

\maketitle

\section{Introduction}\label{sec:intro}

Gravitational collapse in general relativity (GR) generically ends in a singularity~\cite{Penrose:1964wq}. It is widely expected that quantum gravity resolves the singularity, and that the black holes (BHs) of the ``true'' theory possess regular interiors; see, e.g.,~\cite{Lan:2023cvz,Carballo-Rubio:2025fnc} for recent overviews. The regular black hole (RBH) paradigm postulates that an effective metric description remains valid deep inside BH cores. Stationary geometries implementing this paradigm have a long history. Among the most commonly studied are the Bardeen~\cite{Bardeen:1968bh}, Hayward~\cite{Hayward:2005gi}, and Dymnikova~\cite{Dymnikova:1992ux} metrics, which have de Sitter-like cores replacing the Schwarzschild singularity and rapidly approach the Schwarzschild solution outside the event horizon.\footnote{The de Sitter core is not a generic feature, since RBHs with anti-de Sitter or Minkowski cores have also been constructed; see, e.g.,~\cite{Simpson:2021dyo,Panassiti:2025diw,Arrechea:2025nlq}.} 
{All these metrics have the form}
\begin{align}
    \mathrm{d}s^2 = -f(r,M)\mathrm{d}t^2 + \frac{1}{f(r,M)}\mathrm{d}r^2 + r^2\mathrm{d}\Omega^2~,
\end{align}
with $f(r,M)\to1-2M/r$ as $r\to\infty$ and $f(r,M)\sim1-\kappa r^2$, with $\kappa>0$, as $r\to0$. The latter condition ensures that the Kretschmann scalar is finite at the origin.

{The kinematics and basic phenomenology of these geometries are by now well understood}~\cite{Hayward:2005gi,Flachi:2012nv,Carballo-Rubio:2022nuj,Franzin:2023slm}. Their dynamics is not: for most regular metrics, there is no known set of field equations of which they are vacuum solutions, let alone equations capable of describing their formation (without potentially fine-tuning the coupling to specific matter models~\cite{Ayon-Beato:1998hmi, Bronnikov:2000vy, Kudryavcev:2026lbf}). This gap has recently begun to close. {In spherical symmetry, the most general extension of the Einstein equations that is derivable from an action, identically conserved, and of second differential order can be written explicitly}~\cite{Carballo-Rubio:2025ntd, Borissova:2026krh, Borissova:2026wmn}: the construction rests on two-dimensional (2D) Horndeski theory~\cite{Horndeski:1974wa,Kobayashi:2019hrl} as the effective description of the $(t,r)$ sector~\cite{Kunstatter:2015vxa}, and the resulting \emph{master field equations} are parametrised by two free functions $\left(\bm{\alpha}(r,\chi),~\bm{\beta}(r,\chi)\right)$ of the areal radius $r$ and its gradient squared $\chi \coloneqq \nabla_{a}r\,\nabla^{a}r$. 

A Birkhoff theorem holds for the whole class~\cite{Carballo-Rubio:2025ntd}, and the map between theories and their static vacua runs in both directions: given essentially any static metric function $f(r;M)$ with an invertible mass dependence, one can reconstruct the unique pair of theory functions admitting the associated metric as a vacuum solution~\cite{Carballo-Rubio:2025ntd,Boyanov:2025pes,Borissova:2026krh,Bueno:2025zaj}.\footnote{See also~\cite{Bueno:2024dgm,Bueno:2024eig,Bueno:2025zaj,Bueno:2026oyg} for recent advances in dimensions higher than four.} In this precise sense, every RBH now comes with its own gravitational dynamics. 

This very flexibility, however, sharpens the question of selection. If any regular spherically symmetric metric can be promoted to the vacuum of some theory, kinematics alone no longer discriminates between them, and one must ask which of these theories are dynamically sensible. 

To do so, before asking whether collapse in a given theory forms a RBH, one must ask whether the theory can host regular, evolving matter at a regular centre at all, since any formation scenario starts from a regular star or pulse whose centre is an ordinary point of spacetime. This consistency requirement should precede questions about the endpoint of collapse: regularising a vacuum solution may be physically irrelevant if the theory admits no dilute, self-gravitating matter configurations whose evolution can be followed from regular initial data.

In this paper we show that this minimal requirement is surprisingly restrictive. For generic minimally coupled matter, consistency of the field equations at a regular centre forces the theory functions to carry the same definite parities as their GR values: when expressed in terms of $(r,\chi)$ and assumed analytic in $r$ at fixed $\chi$, $\bm{\alpha}$ must be even and $\bm{\beta}$ odd under $r\to-r$. They must also satisfy two centre conditions and a nondegenerate energy-flux law. In the integrable sector, these conditions state that the generalised Misner--Sharp--Hernandez mass is odd across the centre, vanishes as $\mathcal{O}(r^3)$, and contains no point-mass contribution. For theories reconstructed locally from static vacuum families with invertible mass dependence, the parity condition can be read directly from the metric as $f(-r,-M)=f(r,M)$. The Hayward and Dymnikova theories pass this \emph{parity selection rule}; the Bardeen theory fails it. 

We stress that the rule is a necessary centre-consistency condition; it does not assess regularity during dynamical evolution: GR satisfies it and nevertheless evolves generic collapsing matter to singularities. It selects theories in which regular collapse can be posed locally at the centre, without determining the endpoint. An analogous restriction for static stellar spacetimes was identified in~\cite{Arrechea:2026ngi}.

The paper is organised as follows. In section~\ref{sec:framework} we review the master field equations, the potential-function formulation of their vacua, and the coupling to matter. In section~\ref{sec:rule} we define regular centres, derive the selection rule and its companion conditions, and discuss what the rule does and does not assert. Section~\ref{sec:examples} translates the rule into the mass-inversion property of the vacuum metric and works through GR, Hayward, Bardeen and Dymnikova. Section~\ref{sec:family} maps the space of admissible theories. In section~\ref{sec:readings} we develop two complementary readings of the underlying symmetry, and in section~\ref{sec:discussion} we conclude. A short appendix collects the expansions at the centre on which the derivation rests.

\section{Effective field equations for spherical gravity}\label{sec:framework}

We use units of $G=c=1$ and signature $(-+++)$, and we work in four dimensions. Spherically symmetric spacetimes are warped products,
\begin{align}
    g_{\mu\nu}\,\mathrm{d}y^{\mu}\mathrm{d}y^{\nu} = q_{ab}(x)\,\mathrm{d}x^{a}\mathrm{d}x^{b} + r^{2}(x)\,\mathrm{d}\Omega^{2}\,,
\end{align}
with $x^{a}=(t,r)$ the 2D coordinates and the areal radius $r(x)$ treated as a scalar field on the 2D orbit space. 

The construction developed in~\cite{Carballo-Rubio:2025ntd} uses the fact that the most general scalar--tensor action yielding second-order field equations takes the form of a Horndeski theory~\cite{Horndeski:1974wa,Kobayashi:2019hrl}. Taking $q_{ab}(x)$ and $r(x)$ as the tensor--scalar pair, respectively, this theory can therefore describe the dynamics of the warped-product class above. {Varying the $2$D Horndeski action with respect to $q_{ab}(x)$ and $r(x)$} yields the master field equations of Ref.~\cite{Carballo-Rubio:2025ntd}, which deform the spherically symmetric Einstein equations into
\begin{align}
    \mathscr{G}_{\mu\nu}(q,r) = 8\pi\,T_{\mu\nu}\,,
    \label{eq:master}
\end{align}
where $\mathscr{G}_{\mu\nu}$ is the most general spherically symmetric tensor that is symmetric, identically conserved, derivable from an action, and contains at most second derivatives of $q_{ab}$ and $r$. 

Its 2D block is controlled by two free functions $\bm{\alpha}(r,\chi)$ and $\bm{\beta}(r,\chi)$, where
\begin{align}
 \mathscr{G}_{ab}=\frac{\mathscr{E}_{ab}}{r^{2}}\,,
 \end{align} 
with
\begin{align}
    \mathscr{E}_{ab} ={}& \bm{\beta}\,\nabla_{a}\nabla_{b}r - q_{ab}\left(\tfrac{1}{2}\bm{\alpha} + \bm{\beta}\,\Box r\right)\nonumber\\
    &+ \left(\partial_{\chi}\bm{\alpha} - \partial_{r}\bm{\beta}\right)\nabla_{a}r\,\nabla_{b}r\,.
    \label{eq:Eab}
\end{align}

The off-shell Bianchi identity fixes the angular block and carries no independent information once the matter equations of motion hold (see~\cite{Carballo-Rubio:2025ntd} for details). 

GR corresponds to
\begin{align}
    \bm{\alpha}_\textsc{gr}=2\,(1-\chi)\,,\qquad \bm{\beta}_\textsc{gr}=-2r\,,
\end{align}
and we restrict attention to theories that approach these values at large radii, so that the far-field equations are standard in this limit.

In the extended Schwarzschild gauge,
\begin{align}
    q_{ab}\,\mathrm{d}x^{a}\mathrm{d}x^{b} = -n(t,r)^{2}f(t,r)\,\mathrm{d}t^{2} + \frac{\mathrm{d}r^{2}}{f(t,r)}\,,
    \label{eq:gauge}
\end{align}
one has $\chi=f$, and the 2D field equations $\mathscr{E}_{ab}=8\pi r^{2}T_{ab}$ take the compact form~\cite{Carballo-Rubio:2025ntd,Borissova:2026krh}
\begin{align}
    \frac{fn^{2}}{2}\big[\bm{\alpha}+\bm{\beta}\,\partial_{r}f\big] &= 8\pi r^{2}\,T_{tt}\,,\label{eq:tt}\\[2pt]
    \frac{\bm{\beta}\,\partial_{t}f}{2f} &= 8\pi r^{2}\,T_{tr}\,,\label{eq:tr}\\[2pt]
    -\frac{\bm{\alpha}+\bm{\beta}\,\partial_{r}f}{2f} - \frac{\partial_{r}n}{n}\,\bm{\beta} + \bm{\omega} &= 8\pi r^{2}\,T_{rr}\,,\label{eq:rr}
\end{align}
where we defined $\bm{\omega}\coloneqq\partial_{\chi}\bm{\alpha}-\partial_{r}\bm{\beta}$, which vanishes identically in GR.

A feature of this gauge that we shall use repeatedly, and that holds for fully dynamical $f$ and $n$, is that Eqs.~\eqref{eq:tt} and~\eqref{eq:rr} contain no time derivatives whatsoever: all of the time evolution of the geometry is carried by the flux equation~\eqref{eq:tr}. 

When $\bm{\omega}=0$ off shell, both theory functions descend from a single potential function,
\begin{align}
    \bm{\alpha}=\partial_{r}\bm{\Omega}(r,\chi)\,,\qquad
    \bm{\beta}=\partial_{\chi}\bm{\Omega}(r,\chi)\,,
    \label{eq:potential}
\end{align}
and in vacuum {Eq.~\eqref{eq:tt}} integrates to the algebraic statement~\cite{Carballo-Rubio:2025ntd,Boyanov:2025pes}
\begin{align}
    \bm{\Omega}\big(r,\chi=f(r)\big) = 4M\,,
    \label{eq:onshell}
\end{align}
with $M$ the Arnowitt--Deser--Misner mass.\footnote{Our normalisation follows Ref.~\cite{Carballo-Rubio:2025ntd}; the potential of Ref.~\cite{Borissova:2026krh} equals $\bm{\Omega}/2$ and evaluates to $2M$ on shell.} The quantity
\begin{align}
    m(t,r)\coloneqq\frac{\bm{\Omega}(r,\chi)}{4}
    \label{eq:MSmass}
\end{align}
is therefore the generalised Misner--Sharp--Hernandez mass of the theory~\cite{Misner:1964je,Hernandez:1966zia}: it is constant in vacuum, it obeys $\chi=1-2m/r$ in GR by construction, and Eq.~\eqref{eq:tr} is its flux law, $\partial_{t}\bm{\Omega}=16\pi r^{2}f\,T_{tr}$.

The dictionary between theories and their vacua now runs both ways. By the Birkhoff theorem governing theories described by the equations~\eqref{eq:Eab}, the vacuum of a given $(\bm{\alpha},\bm{\beta})$ is a static one-parameter family $f(r;M)$~\cite{Carballo-Rubio:2025ntd}; conversely, given a static metric function with $\partial_{M}f\neq0$, solving $f(r;M) = \chi$ for $M$ yields $\bm{\Omega}(r,\chi)=4M(r,\chi)$ off shell, and {hence fixes the theory locally on the chosen mass branch}. For the Bardeen and Hayward metrics, one has respectively
\begin{align}
    f_\textsc{b} &= 1-\frac{2Mr^{2}}{(r^{2}+\ell^{2})^{3/2}}\,,&
    f_\textsc{h} &= 1-\frac{2Mr^{2}}{r^{3}+2\ell^{2}M}\,,
    \label{eq:BHmetrics}
\end{align}
with $\ell$ the regularization scale. These metric functions lead respectively to
\begin{align}
    \bm{\Omega}_\textsc{b} &= \frac{2(1-\chi)(r^{2}+\ell^{2})^{3/2}}{r^{2}}\,,&
    \bm{\Omega}_\textsc{h} &= \frac{2(1-\chi)\,r^{3}}{r^{2}-\ell^{2}(1-\chi)}\,,
    \label{eq:BHpotentials}
\end{align}
while GR itself is {characterized} by $\bm{\Omega}_\textsc{gr}=2r(1-\chi)$. 

Throughout, we will also use the combination
\begin{align}
    \psi \coloneqq \frac{1-\chi}{r^{2}}\,,
    \label{eq:psi}
\end{align}
which is even in $r$, remains finite at a regular centre, and reduces in GR to $2m/r^{3}$.

Matter enters Eq.~\eqref{eq:master} as any conserved source; only the gravitational side of the equations is deformed. Our reference model is a massless, minimally coupled scalar field, $\phi=\phi(t,r)$ with stress-energy tensor
\begin{align}
    T_{\mu\nu}=\nabla_{\mu}\phi\,\nabla_{\nu}\phi-\tfrac{1}{2}g_{\mu\nu}\,\nabla_{\rho}\phi\,\nabla^{\rho}\phi\,,
\end{align} 
and equation of motion \mbox{$\nabla_{a}(r^{2}q^{ab}\partial_{b}\phi)=0$}. {However, the selection rule derived below uses only the parity and leading-order behaviour of a regular conserved source near the centre; its content is therefore matter agnostic.}

We write $\rho_{c}\coloneqq T_{\mu\nu}u^{\mu}u^{\nu}|_{r=0}$ for the energy density measured by the observer at rest at the centre; {for the scalar field introduced above, this is $\rho_{c}=\tfrac{1}{2}(\dot{\phi}/n)^{2}|_{r=0}$.}

\section{The selection rule}\label{sec:rule}

\subsection{A regular centre is a parity condition}\label{sec:parity}

{We define a centre as regular} when $r=0$ is an ordinary point: nothing diverges, no conical defect, and a local Cartesian chart exists. On the extended radial line, this familiar requirement becomes a parity assignment. A smooth spherically symmetric scalar is a smooth function of $x^{i}x_{i}=r^{2}$, hence even in $r$; the radial component of a regular vector field must vanish at the origin and is odd; and, more generally, a tensor component acquires one sign flip per radial index. 

For the fields of Eq.~\eqref{eq:gauge} and a regular conserved source, this reads
\begin{align}
    &f,\;n,\;\phi,\;\rho\ \text{even}\,;\qquad u^{r}\ \text{odd}\,;\nonumber\\
    &T_{tt},\;T_{rr}\ \text{even}\,;\qquad T_{tr}\ \text{odd}\,,
    \label{eq:parities}
\end{align}
supplemented by elementary flatness, $f(t,0)=1$, which keeps $\psi$ of Eq.~\eqref{eq:psi} finite at the centre. Strict regularity is thus an infinite-order condition concentrated at one point: every odd radial derivative of an even field must vanish at $r=0$, at every instant. 

We stress that this is more restrictive than merely demanding the finiteness of zeroth-order curvature invariants at the centre~\cite{Zhou:2022yio, Antonelli:2025zxh}. 

\subsection{Necessity of the parities}\label{sec:necessity}

We work locally near $r=0$ and consider a theory,  set by a $(\bm{\alpha},\bm{\beta})$ pair, coupled to regular matter. We assume these functions are analytic in $r$ at fixed $\psi$. 
We want to study under which conditions the field equations~\eqref{eq:tt}--\eqref{eq:rr} are consistent with the parity rules described in Eq.~\eqref{eq:parities}.
Each equation relates a combination of theory functions and fields to a source of definite parity. If the two sides carry different parities, the equation splits, forcing both parity projections to vanish separately, yielding an infinite tower of extra conditions at $r=0$ that generic regular data cannot satisfy. The evolution would produce a kink at the centre. A theory in which this happens cannot hold regular matter at a regular centre without fine-tuning. 

Certain evolution schemes, such as those based on double-null coordinates, might bypass these issues by pushing $r=0$ outside the domain of interest in numerical evolution. Yet, the selection rule should not be dismissed, as physical theories should be globally compatible with generic initial data by construction~\cite{Ziprick:2010vb,Barenboim:2025ckx}.

It is instructive to see first how GR passes this test, since the deformed equations must reproduce the same structure. Inserting $\bm{\alpha}_\textsc{gr}=2(1-\chi)$ and $\bm{\beta}_\textsc{gr}=-2r$ into Eqs.~\eqref{eq:tt} and~\eqref{eq:tr}, and trading $f$ for the GR mass function $m=r(1-f)/2$, the two equations collapse to
\begin{align}
    \partial_{r}m = 4\pi r^{2}\,\varepsilon\,,\qquad
    \partial_{t}m = 4\pi r^{2}f\,T_{tr}\,,
    \label{eq:GRtemplate}
\end{align}
where $\varepsilon=T_{tt}/(n^{2}f)$ is the local energy density. Every quantity here carries a definite parity, and the parities match on both sides of each equation. The density $\varepsilon$ is even, so the first equation makes $m$ odd, growing as $m~\simeq~\tfrac{4\pi}{3}\rho_{c}r^{3}$ near a regular centre, and the second balances the odd $\partial_{t}m$ against the odd flux $r^{2}fT_{tr}$. This is the structure a deformation must preserve, and the question is what it demands of $\bm{\alpha}$ and $\bm{\beta}$.

For the general case, we need to determine the behaviour of all fields near the centre. Regularity fixes, at each instant,
\begin{align}
    f &= 1-\psi_{c}r^{2}-\psi_{2}r^{4}-\cdots\,,\qquad
    n = n_{c}\big(1+n_{2}r^{2}+\cdots\big)\,,\nonumber\\
    \phi &= \phi_{c}+\phi_{2}r^{2}+\phi_{4}r^{4}+\cdots\,,
    \label{eq:centreexpansions}
\end{align}
with every coefficient a function of $t$, so that $\psi=\psi_{c}+\psi_{2}r^{2}+\cdots$ and $\partial_{t}f=-\dot{\psi}_{c}r^{2}+\mathcal{O}(r^{4})$. The scalar source then behaves as
\begin{align}
    T_{tt} &= \tfrac{1}{2}\dot{\phi}_{c}^{\,2}+\mathcal{O}(r^{2})\,,\qquad
    T_{tr} = 2\dot{\phi}_{c}\phi_{2}\,r+\mathcal{O}(r^{3})\,,\nonumber\\
    T_{rr} &= \rho_{c}+\mathcal{O}(r^{2})\,,
    \label{eq:matterexpansions}
\end{align}
and the matter coefficients are free data.\footnote{{The equality $T_{rr}|_{r=0}=\rho_{c}$ is specific to the massless scalar; a generic regular source has $T_{rr}=p_{c}+\mathcal{O}(r^{2})$, with finite central pressure $p_{c}$.}}
For generic $\dot\phi_{c}\neq0$, the central density and the leading flux coefficient $2\dot{\phi}_{c}\phi_{2}$ can be varied independently, with either sign for the latter.

Now take the flux equation~\eqref{eq:tr}. The factor $\partial_{t}f/(2f)$ is even, and the source is odd. Splitting $\bm{\beta}$ into even and odd parts in $r$, the even projection of the equation reads
\begin{align}
    \bm{\beta}_{\rm even}(r,\psi)\,\frac{\partial_{t}f}{2f} = 0
    \label{eq:betaeven}
\end{align}
pointwise on every regular solution. Solutions with \mbox{$\partial_{t}f\neq0$} near the centre certainly exist in a healthy theory. If every regular solution had a frozen geometry there, the left-hand side of Eq.~\eqref{eq:tr} would vanish identically while its right-hand side would start being {nonzero} at $\mathcal{O}\left(r^{3}\right)$, so the equation could not be satisfied to arbitrary order. {Hence $\bm{\beta}_{\rm even}$ must vanish on the open set {of} physical solutions; under the analyticity assumption stated above, $\bm{\beta}$ is odd there.}\footnote{At the level of a centre expansion, the field equations force every even-parity Taylor coefficient of $\bm{\beta}$, and every odd-parity Taylor coefficient of $\bm{\alpha}$, to vanish. Smooth nonanalytic terms that are flat at the centre, such as $e^{-1/r^{2}}$ extended by zero at $r=0$, are invisible to all orders and are outside the class considered here. If regular solutions exist in an open neighbourhood where the field equations hold pointwise, the corresponding wrong-parity component is instead excluded directly in that neighbourhood.} With $\bm{\beta}$ odd, the combination $\bm{\beta}\,\partial_{r}f$ in Eq.~\eqref{eq:tt} is even, being a product of two odd factors, and the odd projection of that equation reduces to $\tfrac{1}{2}fn^{2}\bm{\alpha}_{\rm odd}=0$ for an even source. {Thus $\bm{\alpha}$ is even on the same domain.}

Parity is not the whole story. {A solution consistent with the generic initial data~\eqref{eq:matterexpansions} enforces an additional relation between $\bm{\alpha}$ and $\bm{\beta}$ in the limit $r\to0$. Note that
\begin{equation}
    8\pi r^{2}T_{tt}=8\pi r^{2}T_{rr}=\mathcal{O}\left(r^2\right),\quad 8\pi r^{2}T_{tr}=\mathcal{O}\left(r^3\right),
\end{equation}
 are the terms that enter the right-hand side of Eqs.~\mbox{(\ref{eq:tt}--\ref{eq:rr})}.}
The left-hand sides must vanish accordingly. For Eq.~\eqref{eq:tt},
\begin{align}
    \lim_{r\to0}\,\frac{fn^{2}}{2}\big[\bm{\alpha}+\bm{\beta}\,\partial_{r}f\big]
    = \frac{n_{c}^{2}}{2}\lim_{r\to0}\big[\bm{\alpha}+\bm{\beta}\,\partial_{r}f\big] = 0~,
    \label{eq:ttlimit}
\end{align}
at fixed $\psi=\psi_{c}$. 

Suppose for the moment that $\bm{\alpha}$ and $\bm{\beta}$ are finite at the centre:\footnote{We shall lift this assumption later on and find that the field equations impose it.} given that for finite and odd $\bm\beta$ we have $\bm\beta \sim b_1r$, and ${\partial_{r}f \sim -2 r \psi_c}$, then $\bm{\beta}\,\partial_{r}f$ is $\mathcal{O}(r^{2})$, so $\bm{\alpha}$ itself must vanish {at the centre}. 

{A similar analysis reveals that} in Eq.~\eqref{eq:rr} the first two terms are also $\mathcal{O}(r^{2})$, and the limit isolates the last one; hence, taking the $r \to 0$ limit gives
\begin{align}
    \lim_{r\to0}\bm{\omega}(r,\psi_{c}) = 0\,.
    \label{eq:omegalimit}
\end{align}
Both the above statements about the behaviour of $\bm \alpha$ and $\bm \omega$ at the centre hold for every central compactness $\psi_{c}$ compatible with the data. Taken together, they provide what we may call the \emph{centre conditions}
\begin{align}
    \bm{\alpha}\xrightarrow[\ r\to0\ ]{}0\,,\qquad
    \bm{\omega}\xrightarrow[\ r\to0\ ]{}0
    \qquad\text{(fixed $\psi$)}\,.
    \label{eq:centreconditions}
\end{align}
The same limits lift the finiteness assumption made at the start of this discussion, because a left-hand side that blows up cannot match a source that vanishes. 

We now examine possible poles in $\bm{\alpha}$ and $\bm{\beta}$ and show that consistency with generic regular matter excludes them. We work in the class where the theory functions are analytic in $r$ at fixed $\psi$ apart from a pole at the centre of order at most $\bm{\alpha}=\mathcal{O}(1/r)$, $\bm{\beta}=\mathcal{O}(1/r^{2})$. This is the strongest divergence that a potential $\bm{\Omega}$ finite at the centre can generate,\footnote{Through the potential relations \eqref{eq:potential}, which hold when $\bm{\omega}=0$ off shell: $\bm{\alpha}=\partial_{r}\bm{\Omega}$ and $\bm{\beta}=\partial_{\chi}\bm{\Omega}$ evaluated on \mbox{$\bm{\Omega}=\bm{\Omega}_{0}(\psi)+\mathcal{O}(r)$}, with $\partial_{\chi}|_{r}=-r^{-2}\partial_{\psi}$, give the stated orders. Every theory reverse-constructed from a metric is integrable and has such a potential, so the allowed order of irregularity in $\bm{\alpha}$ and $\bm{\beta}$ covers all examples here. For a general theory it is a hypothesis on $(\bm{\alpha},\bm{\beta})$, not a consequence of integrability, and the deeper-pole argument below shows it can be dropped. The centre condition $\bm{\omega}\to0$ derived above holds only at $r=0$ and does not by itself make the theory integrable.} since \mbox{$\bm{\Omega}=\bm{\Omega}_{0}(\psi)+\mathcal{O}(r)$} gives $\bm{\alpha}\supset-2\psi\,\bm{\Omega}_{0}'/r$ and $\bm{\beta}\supset-\bm{\Omega}_{0}'/r^{2}$, and it is exactly what Bardeen saturates;\footnote{This is possible for Bardeen because its $\bm \alpha$ and $\bm \beta$ do not satisfy the parity condition at the centre.} deeper poles, which parity alone would allow, are not allowed as we now argue.

Let us expand $\bm{\alpha}$ and $\bm{\beta}$ around $r=0$ at fixed $\psi$ as 
\begin{align}
    \label{eq:alpha_beta_r0}
    \bm{\alpha}=\sum_{k}a_{k}(\psi)\,r^{k}~,\quad 
    \bm{\beta}=\sum_{k}b_{k}(\psi)\,r^{k}~.
\end{align}
Under the admitted pole order and the requirement of correct parity, the only survivors are a constant term $a_{0}$ and a simple pole $b_{-1}$. Equation~\eqref{eq:ttlimit} alone does not eliminate them, because $\bm{\beta}\,\partial_{r}f\to-2\psi_{c}b_{-1}$ stays finite and the limit~\eqref{eq:ttlimit} only ties the pair together,
\begin{align}
    a_{0}(\psi) = 2\psi\,b_{-1}(\psi)\,.
    \label{eq:a0b1}
\end{align}
The $rr$ equation~\eqref{eq:rr} imposes additional conditions that fix the structure of the poles. Since the field equations differentiate at fixed $\chi$ while the expansions hold $\psi$ fixed, we convert with $\partial_{\chi}|_{r}=-r^{-2}\partial_{\psi}$ and $\partial_{r}|_{\chi}=\partial_{r}|_{\psi}-(2\psi/r)\partial_{\psi}$, which, {using Eq.~\eqref{eq:alpha_beta_r0}}, reduces $\bm{\omega}=\partial_{\chi}\bm{\alpha}-\partial_{r}\bm{\beta}$ into
\begin{align}
    \bm{\omega} = -\sum_{k}a_{k}'\,r^{k-2}-\sum_{k}\big[k\,b_{k}-2\psi\,b_{k}'\big]\,r^{k-1}\,,
    \label{eq:omegaseries}
\end{align}
with $r^{-2}$ coefficient $-a_{0}'+b_{-1}+2\psi b_{-1}'$. Substituting $a_{0}'~=~2b_{-1}~+~2\psi b_{-1}'$ from Eq.~\eqref{eq:a0b1} collapses this coefficient to $-b_{-1}$. The source of Eq.~\eqref{eq:rr} has no $\mathcal{O}\left(r^{-2}\right)$ part, so $b_{-1}=0$, and with it $a_{0}=0$.

Deeper poles are not allowed. Take, for example, the parity-consistent pair $\bm{\alpha}\supset a_{-2}(\psi)\,r^{-2}$ and $\bm{\beta}~\supset~b_{-3}(\psi)\,r^{-3}$. Their contribution to the limit~\eqref{eq:ttlimit} is $a_{-2}-2\psi_{c}b_{-3}$ at order $r^{-2}$, and the tuning $a_{-2}=2\psi\,b_{-3}$ annihilates it: this is the same leading balance of $\bm{\alpha}$ against $\bm{\beta}\,\partial_{r}f$ that produced Eq.~\eqref{eq:a0b1}, now one level deeper. 

The cancellation is undone at the next order by the subleading structure of $f$. From the expansion of $f$ at the centre~\eqref{eq:centreexpansions} the tuned pair leaves the finite residue $-2\psi_{2}\,b_{-3}(\psi_{c})$ in the limit~\eqref{eq:ttlimit}, and $\psi_{2}$ is set by the matter through the central constraint below, not by the theory, so no choice of $b_{-3}$ can remove it. 

{Thus stronger divergences in $\bm{\alpha}$ and $\bm{\beta}$ force otherwise free coefficients of the regular field expansion to vanish. Such theory functions are therefore incompatible with the generic initial data assumed here.}

The $rr$ equation leads to the same result independently: after the tuning, the deepest coefficient of $\bm{\omega}$ collapses to $b_{-3}$ at order $r^{-4}$, an order that no source and no other term of Eq.~\eqref{eq:rr} can meet. The mechanism is uniform in the pole depth. 

{The argument is easily generalizable to deeper poles; thus,} no pole (compatible with parity) of any depth survives. Appendix~\ref{app:centre} records the general coefficient, and shows that in the integrable sector the whole question collapses at once, because there Eq.~\eqref{eq:tt} constrains the mass directly.

The centre conditions~\eqref{eq:centreconditions} therefore hold in the strong sense: finite limits, equal to zero, with no singular parts. In the integrable sector they have a transparent meaning. With $\bm{\Omega}$ odd and Eq.~\eqref{eq:centreconditions}, the mass function \eqref{eq:MSmass} is odd across the centre and vanishes there as $m\sim r^{3}$, which is precisely how the Misner--Sharp--Hernandez mass of any regular distribution behaves in GR. The condition forbids a point mass sitting at $r=0$.

One last limit remains, and it differs in kind from the preceding ones. Those constrained the theory functions; this one yields a dynamical relation, an evolution equation for the central compactness. Dividing Eq.~\eqref{eq:tr} by $r^{3}$ and using $\partial_{t}f=-\dot{\psi}_{c}r^{2}+\cdots$ gives the central flux law,
\begin{align}
    b_{1}(\psi_{c})\,\partial_{t}\psi_{c} = -16\pi\,\big[T_{tr}/r\big]_{r=0}\,,
    \qquad b_{1}\coloneqq\lim_{r\to0}\frac{\bm{\beta}}{r}\,.
    \label{eq:fluxlaw}
\end{align}
The right-hand side is freely specifiable, so $b_{1}$ must not vanish on the range of $\psi$ given by regular data, as in this case the flux law degenerates into a constraint on the matter. With $b_{1}\neq0$ it instead fixes $\partial_{t}\psi_{c}$ from the central flux, the centre-limit counterpart of the GR mass-growth law $\partial_{t}m=4\pi r^{2}fT_{tr}$.

Collecting the results, a theory of this class can host regular, evolving matter at a regular centre only if
\begin{align}
    \ \bm{\alpha}(-r,\psi)=\bm{\alpha}(r,\psi)\,,\qquad
    \bm{\beta}(-r,\psi)=-\bm{\beta}(r,\psi) \,,
    \label{eq:rule}
\end{align}
on the range of $(r,\psi)$ allowed by the data, together with the centre conditions \eqref{eq:centreconditions} and $b_{1}\neq0$. We refer to Eq.~\eqref{eq:rule} as the \emph{parity selection rule}. 

The Hayward and Dymnikova theories considered below satisfy the conditions~\eqref{eq:centreconditions} and~\eqref{eq:rule}, whereas Bardeen fails both of them. The companion conditions also have independent content: parity-correct theory functions with sufficiently deep centre poles are excluded by the analysis above. Conversely, when the rule and the centre conditions hold, no parity or formal centre-expansion obstruction remains. Each equation carries a single parity, and the expansion of the system around $r=0$ closes order by order, with every metric coefficient determined from the matter.

The leading instance generalises the first of Eqs.~\eqref{eq:GRtemplate}: the $\mathcal{O}\left(r^{2}\right)$ part of Eq.~\eqref{eq:tt} is the central constraint,
\begin{align}
    a_{2}(\psi_{c})-2\,b_{1}(\psi_{c})\,\psi_{c}=16\pi\rho_{c}\,,
    \qquad a_{2}\coloneqq\lim_{r\to0}\frac{\bm{\alpha}}{r^{2}}\,,
    \label{eq:centralconstraint}
\end{align}
which fixes the central compactness from the central density. 

In GR, $\bm{\alpha}_\textsc{gr}=2(1-\chi)=2\psi r^{2}$ gives $a_{2}=2\psi_{c}$ and $\bm{\beta}_\textsc{gr}=-2r$ gives $b_{1}=-2$, so the left-hand side is $6\psi_{c}$ and the constraint becomes $\psi_{c}=\tfrac{8\pi}{3}\rho_{c}$, equivalently the familiar $m\simeq\tfrac{4\pi}{3}\rho_{c}r^{3}$ near the centre. In practical terms, the rule is equivalent to the statement that the standard order-by-order construction of regular initial data at the centre applies to generic matter, exactly as it does in GR.

For exactly static configurations, Eq.~\eqref{eq:tr} is vacuous; the wrong-parity projections become constraints, and specially tuned static profiles might satisfy them. A theory that fails the rule may therefore admit isolated, fine-tuned stars in equilibrium, as in~\cite{Arrechea:2026ngi}, but it cannot evolve generic matter through its centre. And the rule constrains the theory functions, at fixed $\psi$ up to the largest central compactness available by the data, which Eq.~\eqref{eq:centralconstraint} ties to the central density.

\subsection{What the rule does and does not assert}\label{sec:scope}

The parity selection rule is a consistency condition on the initial-value problem at the centre. It is necessary for any formation scenario that starts from regular data, since such data contain a regular centre by assumption. It is not a formation result. GR passes the rule; nonetheless, generic scalar collapse produces singularities: parity consistency protects the smoothness structure of the centre for finite field values, and places no bound on the fields themselves. 

A centre that collapses parity-symmetrically, with $\rho_{c}(t)\to\infty$ in finite time, respects every statement above. The singularity is a breakdown of the evolution, the curvature diverging in finite time, and not an inconsistency of the field equations at $r=0$. Likewise, nothing here addresses the hyperbolicity of the system at large field values, a known point of concern in Horndeski-type theories~\cite{Papallo:2017qvl, Ripley:2022cdh}, nor the endpoint the dynamics selects: the Birkhoff theorem fixes the vacuum exterior of a collapsed region to be the theory's static solution, but whether the interior can relax to the regular vacuum core is a dynamical question that the rule leaves entirely open.

What the rule does deliver, besides cutting the theory space, is a clean diagnostic: in an admissible theory the centre analysis is consistent at every amplitude, so if an evolution breaks down, the breakdown is physics rather than an inconsistency in prescribing regular data.

\section{Reading the rule off the metric}\label{sec:examples}

\subsection{Mass inversion}\label{sec:massinversion}

For integrable theories the rule condenses into a statement about $\bm{\Omega}$: since $\bm{\alpha}=\partial_{r}\bm{\Omega}$ is even and $\bm{\beta}=\partial_{\chi}\bm{\Omega}$ odd, the potential itself must be odd in $r$ once its additive constant is fixed. Through the reverse construction, this becomes a property one can read directly off the vacuum metric. 

Suppose $f(r;M)$ has an invertible mass dependence, and let $M(r,f)$ denote the inversion, so that $\bm{\Omega}=4M(r,\chi)$. If the metric function satisfies
\begin{align}
    f(-r,-M) = f(r,M)\,,
    \label{eq:massinversion}
\end{align}
then evaluating it at $(-r,-M(r,f_{0}))$ returns $f_{0}$, and uniqueness of the inversion gives $M(-r,f_{0})=-M(r,f_{0})$: the potential is odd. Running the argument backwards shows the two properties are equivalent. 

The selection rule for reverse-constructed theories is therefore the invariance of the vacuum metric under a simultaneous inversion of radius and mass, Eq.~\eqref{eq:massinversion}. Schwarzschild obeys it, since $1-2M/r$ depends on radius and mass only through their ratio. Note also the immediate negative consequence: if $f$ is even in $r$ at fixed $M$, then $M(r,f)$ is even, and the potential cannot be odd, so any such theory fails the rule.

\subsection{Examples}\label{sec:fourexamples}
Let us illustrate how the selection rule applies to theories with RBHs as vacuum solutions. GR passes the test, since $\bm{\Omega}_\textsc{gr}=2r(1-\chi)=2r^{3}\psi$ is odd, while $\bm{\alpha}_\textsc{gr}$ is even and $\bm{\beta}_\textsc{gr}$ odd, with $b_{1}=-2$. We focus on the most widely used examples.

{\bf Hayward RBH}: From Eqs.~\eqref{eq:BHmetrics} and~\eqref{eq:BHpotentials}, we can easily derive \mbox{$\bm{\Omega}_\textsc{h}=2r^{3}\psi/(1-\ell^{2}\psi)$}, which is odd, and explicitly check that
\begin{align}
    \bm{\alpha}_\textsc{h} &= \frac{2(1-\chi)r^{2}\big[r^{2}-3\ell^{2}(1-\chi)\big]}{\big[r^{2}-\ell^{2}(1-\chi)\big]^{2}}\,,\nonumber\\
    \bm{\beta}_\textsc{h} &= -\frac{2r^{5}}{\big[r^{2}-\ell^{2}(1-\chi)\big]^{2}}\,,
    \label{eq:haywardab}
\end{align}
are respectively even and odd, with $b_{1}=-2/(1-\ell^{2}\psi)^{2}\neq0$ and $\bm{\alpha}_\textsc{h}=\mathcal{O}(r^{2})$ at fixed $\psi$: all conditions hold, and $f_\textsc{h}(-r,-M)=f_\textsc{h}(r,M)$ is immediate from Eq.~\eqref{eq:BHmetrics}. The theory functions possess a pole on the surface determined by the condition $\ell^{2}\psi=1$: this is the limiting density of the theory, approached on shell only as $r\to0$ in the static solution, or as $\rho_{c}\to\infty$ through the central constraint, which for Hayward reads
$\psi_{c}=8\pi\rho_{c}/\left(3+8\pi\ell^{2}\rho_{c}\right)$. {The limiting static core deserves a qualification. At fixed nonzero $M$,
$f_\textsc{h}=1-r^{2}/\ell^{2}+r^{5}/(2\ell^{4}M)+\mathcal{O}(r^{8})$, so it is analytic as a function of the coordinate $r$. Still, it is not even and therefore does not define a $C^{\infty}$ spherically symmetric scalar on a Cartesian neighbourhood in the strict sense of Sec.~\ref{sec:parity}. The selection rule states that the Hayward \emph{theory} can support generic regular matter centres for finite $\psi<1/\ell^{2}$. } Since the matter distribution always covers the centre and is never truly in vacuum (except Minkowski), such non-regular behaviour of the Hayward solution at $r=0$ is never problematic.

{\bf Bardeen RBH}: $\bm{\Omega}_\textsc{b}=2\psi\,(r^{2}+\ell^{2})^{3/2}$ is even in $r$: the parities come out inverted, {with}
\begin{align}
    \bm{\alpha}_\textsc{b} &= \frac{2(1-\chi)\sqrt{r^{2}+\ell^{2}}\,\big(r^{2}-2\ell^{2}\big)}{r^{3}}\,,\nonumber\\
    \bm{\beta}_\textsc{b} &= -\frac{2\,(r^{2}+\ell^{2})^{3/2}}{r^{2}}\,,
    \label{eq:bardeenab}
\end{align}
respectively odd and even. {These are the exact opposite of the parities required by Eq.~\eqref{eq:rule}.} The failure is visible already in Eq.~\eqref{eq:BHmetrics}: $f_\textsc{b}$ is even in $r$ at fixed $M$, the situation excluded at the end of Sec.~\ref{sec:massinversion}. It is compounded at the centre, where at fixed $\psi$ one finds $\bm{\alpha}_\textsc{b}\simeq-4\ell^{3}\psi/r$ and $\bm{\beta}_\textsc{b}\simeq-2\ell^{3}/r^{2}$: the theory functions diverge {at} the centre at any nonzero central density.  {Within the analytic class considered here, the reverse-constructed Bardeen theory therefore cannot support generic evolving matter at a regular centre.}

{This result is counterintuitive from the standpoint of curvature regularity alone: $f_\textsc{b}$ is analytic in $r^{2}$ and defines a smoother Cartesian centre than the fixed-$M$ Hayward function, whose expansion contains an $r^{5}$ term. Kinematic differentiability of a chosen vacuum metric and dynamical centre consistency of its reverse-constructed theory (in the presence of matter) are distinct requirements, and the selection rule tests the latter.}

The companion centre conditions can exclude theories that satisfy parity. For example, consider~\cite{Arrechea:2026ngi}
\begin{eqnarray}\label{eq:Betaodd}
    {\bm{\beta}}_{\rm odd}=-2r-\frac{2\ell^2}{\sqrt{r^2+\ell^2}}\left(\frac{\ell}{r}\right)^{2n+1},\quad  n\geq1
\end{eqnarray}
with $\bm{\alpha}(r,\chi)=-(1-\chi)\,\mathrm{d}\bm{\beta}(r)/\mathrm{d}r$. Here $\bm{\beta}$ is odd and $\bm{\alpha}$ even, but near the centre $\bm{\beta}=\mathcal{O}(r^{-2n-1})$ and, at fixed $\psi$, $\bm{\alpha}=\mathcal{O}(r^{-2n})$. The parity rule is satisfied, while the centre conditions fail. This provides an explicit case in which dynamical centre consistency is more restrictive than the parity test alone and than the corresponding static analysis.

{\bf Dymnikova RBH}: The metric function $f_\textsc{d}=1-\tfrac{2M}{r}\big(1-e^{-r^{3}/(2M\ell^{2})}\big)$ requires more care, because its mass inversion is not elementary. Setting $u\coloneqq r^{3}/(2M\ell^{2})$ and $x\coloneqq\ell^{2}\psi$, the on-shell relation {reduces} to $x=g(u)$ with $g(u)=(1-e^{-u})/u$, strictly decreasing from $1$ to $0$ on $u>0$. The inversion therefore exists and is unique for all $M>0$, and it is explicit in terms of the principal Lambert branch~\cite{Konoplya:2024kih},
\begin{align}
    \bm{\Omega}_\textsc{d} = \frac{2r^{3}\psi}{1+\ell^{2}\psi\,W_{0}\!\left(-\dfrac{e^{-1/(\ell^{2}\psi)}}{\ell^{2}\psi}\right)}\,,
    \label{eq:dymnikova}
\end{align}
which stays on $W_{0}$ throughout the physical branch and is odd in $r$, since $r^{3}$ is odd and $\psi$ even: Dymnikova passes the rule. Its potential shares the simple pole of Hayward's on the limiting-density surface $\ell^{2}\psi=1$, with the same interpretation, and the secondary Lambert branch $W_{-1}$ is reached only upon continuation to $M<0$; the branch structure belongs to the representation, not to the theory, because $g$ is analytic with $g'(0)\neq0$.

\section{The space of admissible theories}\label{sec:family}

The rule is restrictive, but it is far from pinning down a unique theory, and it is instructive to see how large the admissible space is. Any potential that is odd in $r$, has GR asymptotics, respects the centre conditions~\eqref{eq:centreconditions}, and yields $b_{1}\neq0$ on the relevant domain defines an admissible theory at the level of the centre expansion. A convenient one-function family is
\begin{align}
    \bm{\Omega} = 2r^{3}\,\mathbb{h}(\psi)\,,\qquad
    \mathbb{h}(0)=0\,,\quad \mathbb{h}'(0)=1\,,\quad \mathbb{h}'>0\,,
    \label{eq:family}
\end{align}
which is odd for any profile $\mathbb{h}$, reduces to GR for $\mathbb{h}(\psi)=\psi$, and has $\bm{\beta}=-2r\,\mathbb{h}'(\psi)$, hence $b_{1}=-2\mathbb{h}'\neq0$ automatically. On shell, Eq.~\eqref{eq:onshell}, {takes in this case the form}
\begin{align}
    \mathbb{h}\big(\psi(r)\big) = \frac{2M}{r^{3}}\,,
    \label{eq:onshellh}
\end{align}
and the central constraint takes the compact form $3\,\mathbb{h}(\psi_{c})=8\pi\rho_{c}$, with GR recovered as the case $\mathbb{h}(\psi)=\psi$.

Equation~\eqref{eq:onshellh} makes the origin of de Sitter cores in this family transparent. As $r\to0$ the right-hand side diverges, so $\psi(r)$ is pushed towards the upper end of the range of $\mathbb{h}$. If $\mathbb{h}$ grows without bound only as $\psi\to\psi_{\star}$ for some finite $\psi_{\star}$, i.e., if the profile has a vertical asymptote, then $\psi(r)\to\psi_{\star}$. Since $f=1-\psi r^{2}$ is just the definition of $\psi$ rearranged, the core is de Sitter,
\begin{align}
    f(r) = 1-\psi_{\star}\,r^{2}+\dots\,,
\end{align}
with limiting density $\psi_{\star}$ a fixed property of the theory, the same for all masses, thereby realising a limiting-curvature behaviour in the spirit of Markov's hypothesis~\cite{Markov:1982rcm}. The correction is controlled by how fast $\psi$ reaches $\psi_{\star}$: for a simple pole $\mathbb{h}\simeq c/(\psi_{\star}-\psi)$, inverting Eq.~\eqref{eq:onshellh} gives $\psi_{\star}-\psi(r)\simeq c\,r^{3}/2M$, so $f=1-\psi_{\star}r^{2}+\mathcal{O}(r^{5})$. 

If instead $\mathbb{h}$ has no finite asymptote, as in GR where $\mathbb{h}(\psi)=\psi$, then Eq.~\eqref{eq:onshellh} reads $\psi(r)=2M/r^{3}$, which diverges as $r\to0$: there is no limiting density and no $r^{2}$ core, and $f=1-\psi r^{2}=1-2M/r$ keeps the Schwarzschild singularity. We write this degenerate case as $\psi_{\star}=\infty$. 

Horizons sit at $\chi=0$, i.e., $\psi=1/r^{2}$, so a horizon at radius $r_{h}$ has $\psi_{h}=1/r_{h}^{2}$, and Eq.~\eqref{eq:onshellh} returns the corresponding mass at such radius,
\begin{align}
    M(r_{h}) = \tfrac12\,r_{h}^{3}\,\mathbb{h}\big(1/r_{h}^{2}\big)\,,
    \label{eq:horizoncurve}
\end{align}
defined for $r_{h}>\psi_{\star}^{-1/2}$, so that $\psi_{h}$ stays below the pole. This curve approaches infinity at both ends, for different reasons. 

As $r_{h}\to\infty$ the argument $1/r_{h}^{2}\to0$ and $\mathbb{h}(1/r_{h}^{2})\to\mathbb{h}(0)=0$; the growth is not in $\mathbb{h}$ but in the prefactor, since $\mathbb{h}'(0)=1$ makes $\mathbb{h}(1/r_{h}^{2})$ vanish only as $1/r_{h}^{2}$, which $r_{h}^{3}$ outgrows, leaving $M\simeq r_{h}/2\to\infty$, the Schwarzschild relation for large BHs. 

As $r_{h}\to\psi_{\star}^{-1/2}$ from above, the argument approaches the pole $\psi_{\star}$; $\mathbb{h}$ diverges at a finite prefactor, and $M\to\infty$ again: a horizon pushed towards the core scale leads to unbounded mass. A continuous, positive function that diverges at both ends of its interval attains a positive minimum $M_{\star}$ at some $r_{\star}$ in between.

Solving $M(r_{h})=M$ then gives a pair of inner and outer horizons {for $M>M_{\star}$}, a single extremal horizon at $M=M_{\star}$, and none for $M<M_{\star}$. Thus the static vacuum family is horizonless below $M_{\star}$. The regular core produces this well-known feature: in GR, where $\psi_{\star}=\infty$, the curve $M(r_{h})=r_{h}/2$ is monotonic, and horizons exist at every mass. For Hayward, $\mathbb{h}=\psi/(1-\ell^{2}\psi)$ gives $M(r_{h})=r_{h}^{3}/[2(r_{h}^{2}-\ell^{2})]$, minimised at the extremal radius $r_{\star}=\sqrt{3}\,\ell$ with $M_{\star}=3\sqrt{3}\ell/4$.

{Notice also that the Hayward geometry is singled out by rationality. A M\"obius (or linear-fractional) profile is a ratio of two linear functions, $\mathbb{h}(\psi)=(a\psi+b)/(c\psi+d)$, and is the simplest nonpolynomial rational profile.} The normalisation of Eq.~\eqref{eq:family} almost fixes it: $\mathbb{h}(0)=0$ forces $b=0$, $\mathbb{h}'(0)=1$ forces $a=d$, and a pole at the limiting density $\psi_{\star}=1/\ell^{2}$ forces $c=-\ell^{2}d$; the common scale cancels and leaves $\mathbb{h}=\psi/(1-\ell^{2}\psi)$, which is Hayward's. 
 
{Thus Hayward is the unique M\"obius member of the one-function family~\eqref{eq:family}.} This has a sharper form on the vacuum metric. Passing from $\mathbb{h}$ to $f$ means inverting the on-shell relation \eqref{eq:onshellh}, and a rational function is inverted by another rational function only when it is M\"obius: a profile of higher degree is several-to-one, so its inverse is multivalued and carries a root or a transcendental function. Since $\psi=(1-f)/r^{2}$ and $\mathbb{h}(\psi)=2M/r^{3}$, a rational inverse is exactly what makes $f$ rational in $r$ and $M$. 

For Hayward the inversion is immediate: $\psi/(1-\ell^{2}\psi)=2M/r^{3}$ gives $\psi=2M/(r^{3}+2\ell^{2}M)$, and $f=1-\psi r^{2}$ is $f_\textsc{h}$ of Eq.~\eqref{eq:BHmetrics}. Thus \emph{Hayward is the unique member of the family whose metric function is rational in $r$ and $M$}. For a non-M\"obius rational profile $\mathbb{h}$, the metric function is generally algebraic and branch-dependent; more general profiles can lead to transcendental inversions. Dymnikova provides the Lambert-$W$ example in Eq.~\eqref{eq:dymnikova}.

The rule thus selects a structure rather than a metric. The one-parameter deformation $\mathbb{h}_{p}(\psi)=\psi\,(1-\ell^{2}\psi)^{-p}$ satisfies every condition for all $p>0$, with the same limiting density $1/\ell^{2}$, and {at $p=2$ the solution is explicitly:}
\begin{align}
    f_{p=2}(r) = 1-\frac{4M/r}{\,1+\dfrac{4M\ell^{2}}{r^{3}}+\sqrt{1+\dfrac{8M\ell^{2}}{r^{3}}}\,}\,,
    \label{eq:p2metric}
\end{align}
a new Hayward-like RBH with a de Sitter core and a manifest GR limit for $\ell\to0$. 

{Beyond the one-function family~\eqref{eq:family}, the space is larger still. For example,} $\bm{\Omega}=2r^{3}\psi/(1-\ell^{2}\psi)+2\lambda\ell^{4}r^{5}\psi^{3}$ is odd with unaltered asymptotics, so the admissible space does not reduce to Eq.~\eqref{eq:family}. Whether additional physical requirements, such as curvature uniform boundedness in the mass or thermodynamic consistency, single out Hayward within this space is an interesting question that we leave open.

\section{Two readings of the symmetry}\label{sec:readings}

The invariance \eqref{eq:massinversion} admits two physical readings, and we develop both, since each illuminates a different aspect of the rule.

\subsection{An orientation-reversal analogy}
\label{sec:cpt}

{In the integrable sector, we interpret the parity rule in terms of discrete operations.} The reflection $r\to-r$ about the centre worldline is the parity operation P of the reduced 2D theory, and the selection rule states that the quasi-local mass function is P-odd,
\begin{align}
    m(t,-r) = -\,m(t,r)\,,
    \label{eq:modd}
\end{align}
just as in GR. Now, $m$ evaluated in vacuum is the conserved charge associated with the time-translation Killing field, the on-shell value of $\bm{\Omega}/4$. 

{A charge linear in its generator flips sign when the time orientation assigned to that generator is reversed: declaring the past-directed generator to be future-directed relabels $M\to-M$. In this restricted sense, orientation reversal implements mass inversion, and Eq.~\eqref{eq:massinversion} resembles an invariance under P combined with an orientation-reversing T operation; charge conjugation acts trivially for the neutral matter considered here. The convention-independent statement is only that the vacuum \emph{family} is covariant under the simultaneous transformation $(r,M)\to(-r,-M)$.}

The T operation used here reverses the choice of time orientation; it is not the coordinate isometry $t\to-t$ at fixed labels, under which a static configuration retains its mass. Because the mass is the Noether charge of the time-translation generator, it changes sign under the former relabelling {but remains invariant} under the latter. The sharp content remains Eq.~\eqref{eq:modd}; no independent CPT statement is being derived.

{Lastly, recent results indicate that adding charge to RBHs does not always preserve regularity~\cite{Carballo-Rubio:2026mvj}. The relation between those results and the parity selection rule remains to be explored.}

\subsection{Mass inversion as centre crossing}\label{sec:crossing}

The second interpretation uses the sign of the areal radius as a geometric bookkeeping tool, extending it through the centre to negative values.  
This is equivalent to continuing along a radial curve with an antipodal angular identification. In this case, both $f_\textsc{b}$ and $f_\textsc{h}$ admit analytic continuations as one-dimensional functions of signed $r$ at fixed $M$. However, only $f_\textsc{b}$ is even and compatible with the strict Cartesian smoothness criterion of Sec.~\ref{sec:parity}. Accordingly, this continuation of the vacuum family should not be conflated with the existence of a single $C^{\infty}$ spacetime across the Hayward limiting core. 

On the negative axis, written as $r'=-r>0$, the continued metric function is $f(-r';M)$. Reading as $f(-r';M)=f(r';-M)$, the mass-inversion identity \eqref{eq:massinversion} rewrites this as the same solution at the opposite mass. For Bardeen, $f_\textsc{b}$ is even in $r$, so $f_\textsc{b}(-r';M)=f_\textsc{b}(r';M)$: {the signed continuation is mirror symmetric at fixed $M$}. For Hayward, the corresponding continuation within the vacuum family is mass inverted,
\begin{align}
    f_\textsc{h}(r';-M) = 1+\frac{2Mr'^{2}}{r'^{3}-2\ell^{2}M}\,,
    \label{eq:haywardinverted}
\end{align}
whose denominator vanishes at $r'^{3}=2\ell^{2}M$; there $\psi=(1-f)/r^{2}$ diverges. The global structure of such extensions, and the geodesic completeness of some RBHs more broadly, is analysed in Refs.~\cite{Zhou:2022yio, Antonelli:2025zxh}.

The dynamical content of the rule is as follows: a regular centre is a point that matter crosses. Radial geodesics and wave fronts pass through $r=0$ and re-emerge on the antipodal side, and the parity assignment \eqref{eq:parities} is nothing but the bookkeeping of that crossing. For the field equations to transport regular matter through the centre without generating a kink, they must be covariant under the crossing, and that is precisely the selection rule. 

Theories built on mirror-symmetric potentials fail because their field equations cannot transport parity-definite generic matter through the origin, irrespective of the curvature regularity of a particular vacuum metric. Thus, at the level of the vacuum family, compatibility with dynamical centre parity favours the mass-inversion covariance over fixed-mass mirror symmetry. Whether either signed continuation represents a physical maximal extension, and whether that extension is geodesically complete, are separate global questions. The rule itself is agnostic about them: it constrains the theory locally at $r=0$, already for an ordinary star.

In summary, the two interpretations are complementary but secondary to the mathematical result. The orientation-reversal analogy packages the invariance as a statement about charges and conventions, whereas the crossing interpretation relates it to the transport of matter through the origin. Both refer to the same identity, Eq.~\eqref{eq:massinversion}.

\section{Discussion}\label{sec:discussion}

Within the master field equations for spherically symmetric gravity, we have derived a necessary local condition for an analytic theory to describe generic regular matter at a regular centre: the theory functions must carry GR-like parities, with $\bm{\alpha}$ even and $\bm{\beta}$ odd in $r$ at fixed $\psi$, the centre conditions~\eqref{eq:centreconditions} must hold, and the central flux law must remain nondegenerate. In the integrable sector, these requirements imply that the quasilocal mass is odd and vanishes as $r^{3}$ with no point-mass contribution.

For reverse-constructed theories on an invertible mass branch, the parity rule is equivalent to covariance of the vacuum family under simultaneous inversion of radius and mass. The theories associated with the Hayward and Dymnikova metrics pass, whereas Bardeen fails. This comparison separates Cartesian differentiability of a particular static core from dynamical centre consistency of the associated theory. The admissible space is infinite. An important subclass is organised around profiles $\mathbb{h}(\psi)$ whose poles fix a mass-independent limiting density, with Hayward distinguished as the unique member whose metric function is rational in $r$ and $M$. The mass-inversion symmetry also admits orientation-reversal and centre-crossing interpretations, although the identity~\eqref{eq:massinversion} is the primary result. The rule does not determine the outcome of collapse, hence three dynamical questions appear particularly pressing.

First, the fate of matter. For an admissible theory, no parity or formal centre-expansion obstruction prevents posing regular collapse data. The remaining question is dynamical: whether focusing matter continues to collapse, bounces, or settles into a regular core.
A systematic study of this process, including where the evolution breaks down and whether an inner horizon forms first, is worth pursuing. 

Second, the endpoint problem. In the Hayward-like BH branch, the limiting static core is a vacuum configuration behind an inner horizon and lies at the boundary of the regular finite-$\psi$ domain. Trapped matter that continues to focus has no evident route to that state. Whether its formation requires a bounce, as in Hayward's original proposal~\cite{Hayward:2005gi}, and what is the eventual fate of collapsing matter in these theories deserve a dedicated analysis.

Third, the parity selection rule can also serve as a design principle. Imposing Eq.~\eqref{eq:massinversion} at the model-building stage provides an immediate necessary filter, although the centre conditions, nondegenerate flux law, and independent well-posedness requirements must still be checked. We plan to investigate these questions in future work.

\acknowledgments
The authors wish to thank R.~Carballo--Rubio and J.~Borissova for illuminating discussions.
FT is supported by the INFN {postdoctoral} research agreement No. 27076.

\appendix

\section{Expansions at the centre}\label{app:centre}
We collect the expansions behind {Sec.~\ref{sec:rule}}. Regularity ({Sec.~\ref{sec:parity}}) fixes, at each instant,
\begin{align}
    f &= 1-\psi_{c}r^{2}-\psi_{2}r^{4}-\cdots\,,\quad
    n = n_{c}\big(1+n_{2}r^{2}+\cdots\big)\,,\nonumber\\
    \phi &= \phi_{c}+\phi_{2}r^{2}+\phi_{4}r^{4}+\cdots\,,
\end{align}
with all coefficients functions of $t$, so that $\psi(t,r)=\psi_{c}+\psi_{2}r^{2}+\cdots$ and $\partial_{t}f=-\dot{\psi}_{c}r^{2}+\mathcal{O}(r^{4})$. For the scalar source,
\begin{align}
    T_{tt} &= \tfrac{1}{2}\dot{\phi}_{c}^{2}+\big[\dot{\phi}_{c}\dot{\phi}_{2}+2n_{c}^{2}\phi_{2}^{2}\big]r^{2}+\cdots\,,\nonumber\\
    T_{tr} &= 2\dot{\phi}_{c}\phi_{2}\,r+\cdots\,,\qquad
    T_{rr} = \rho_{c}+\mathcal{O}(r^{2})\,,
\end{align}
with $\rho_{c}=\dot{\phi}_{c}^{2}/(2n_{c}^{2})$; {for the scalar, the coefficients of the matter expansion are free data and, for generic $\dot\phi_{c}\neq0$, the central density and leading flux coefficient $2\dot{\phi}_{c}\phi_{2}$ can be varied independently.} Writing $\bm{\alpha}=\sum_{k}a_{k}(\psi)r^{k}$ and $\bm{\beta}=\sum_{k}b_{k}(\psi)r^{k}$ at fixed $\psi$, the chart identity
\begin{align}
    \bm{\omega} = -\sum_{k}a_{k}'\,r^{k-2}-\sum_{k}\big[k\,b_{k}-2\psi\,b_{k}'\big]r^{k-1}
    \label{eq:omegachart}
\end{align}
gives in particular $\bm{\omega}_{0}=-a_{2}'-b_{1}+2\psi b_{1}'$ for regular parities.

The elimination of singular coefficients invoked in Sec.~\ref{sec:necessity} proceeds as follows. The even part of Eq.~\eqref{eq:tr} and the odd part of Eq.~\eqref{eq:tt} remove all even coefficients of $\bm{\beta}$ and odd coefficients of $\bm{\alpha}$, poles included. The limit of Eq.~\eqref{eq:tt} then leaves the single relation $a_{0}=2\psi\,b_{-1}$ between the surviving singular pieces, and substituting it into the $r^{-2}$ coefficient of Eq.~\eqref{eq:rr}, which by Eq.~\eqref{eq:omegachart} reads $-a_{0}'+b_{-1}+2\psi b_{-1}'$, collapses that coefficient to $-b_{-1}$: the source has no $r^{-2}$ component, so $b_{-1}=0$, and with it $a_{0}=0$. The same computation at greater depth excludes every parity-consistent pole: a pair tuned as $a_{-2m}=2\psi\,b_{-2m-1}$ to cancel the order-$r^{-2m}$ part of Eq.~\eqref{eq:tt} leaves the residue $-2\psi_{2}\,b_{-2m-1}(\psi_{c})$ two orders up, where $\psi_{2}$ is free matter data, and by Eq.~\eqref{eq:omegachart} the matching coefficient of $\bm{\omega}$ collapses to $(2m-1)\,b_{-2m-1}$ at order $r^{-2m-2}$, which {the negative-power part of the $rr$ equation} sends to zero; the elimination above is the case $m=0$. The finite limit of Eq.~\eqref{eq:rr} is then $\bm{\omega}_{0}=0$, which is the second of the centre conditions~\eqref{eq:centreconditions} and, in the integrable sector, is equivalent to the absence of a point mass, and no restriction on the pole order is needed at all. Since $\bm{\alpha}+\bm{\beta}\,\partial_{r}f=\mathrm{d}\,\bm{\Omega}(r,f)/\mathrm{d}r=4\,\partial_{r}m$, Eq.~\eqref{eq:tt} forces the composite mass to be regular directly, the deepest Laurent coefficient of $\bm{\Omega}$ must vanish for every reachable $\psi_{c}$, and the induction ascends through the tail order-by-order, so any mass structure pinned to the origin is excluded.

Once the rule and the centre conditions hold, the system closes order-by-order. Using $\bm{\omega}_{0}=0$ in the form $a_{2}'=-b_{1}+2\psi b_{1}'$, one finds that the order-$r^{2k+2}$ part of Eq.~\eqref{eq:tt} contains the newest metric coefficient $\psi_{2k}$ with the universal factor
\begin{align}
    a_{2}'-2\psi_{c}b_{1}'-(2k+2)\,b_{1} = -(2k+3)\,b_{1}\,,
\end{align}
the order-$r^{2k}$ part of Eq.~\eqref{eq:rr} contains $n_{2k}$ with factor $-2k\,b_{1}$, and the order-$r^{2k+3}$ part of Eq.~\eqref{eq:tr} contains $\dot{\psi}_{2k}$ with factor $-b_{1}/2$. The three hierarchies interleave triangularly, each order determining exactly one new coefficient from the matter expansion, so for $b_{1}\neq0$ the centre analysis closes at all orders, exactly as in GR, where the factors reduce to $2(2k+3)$, $4k$ and $1$. The leading instances are the central constraint \eqref{eq:centralconstraint}, the flux law \eqref{eq:fluxlaw}, and the slicing condition $n_{2}=[\omega_{2}-8\pi(\rho_{c}+p_{c})]/(2b_{1})$ with $p_{c}$ the central radial pressure, whose GR value is $n_{2}=4\pi\rho_{c}$ for the scalar.

\bibliography{biblio}

\end{document}